\documentclass[12pt]{article}
\usepackage{graphicx}
\pdfoutput=1

\newcommand\pubnumber{SNSN-323-63}
\newcommand\pubdate{\today}

\def\institute{Deutsches Elektronen-Synchroton (DESY)\\
Notkestr. 85, 22607 Hamburg, GERMANY}

\def\Title#1{\begin{center} {\Large #1 } \end{center}}
\def\Author#1{\begin{center}{ \sc #1} \end{center}}
\def\Address#1{\begin{center}{ \it #1} \end{center}}

\newcommand\pubblock{\rightline{\begin{tabular}{l} \pubnumber\\
         \pubdate  \end{tabular}}}
\newenvironment{Abstract}{\begin{quotation}  }{\end{quotation}}
\newenvironment{Presented}{\begin{quotation} \begin{center} 
             PRESENTED AT\end{center}\bigskip 
      \begin{center}\begin{large}}{\end{large}\end{center} \end{quotation}}

\def\beq{\begin{equation}}
\def\eeq#1{\label{#1}\end{equation}}
\def\eeqn{\end{equation}}

\def\beqa{\begin{eqnarray}}
\def\eeqa#1{\label{#1}\end{eqnarray}}
\def\eeqan{\end{eqnarray}}

\let\bar=\overbar

\def\Dslash{\not{\hbox{\kern-4pt $D$}}}
\def\dslash{\not{\hbox{\kern-2pt $\del$}}}

\def\msb{{\bar{\ssstyle M \kern -1pt S}}}

\begin{document}
\begin{titlepage}
\pubblock

\vfill
\Title{{${\rm t\bar{t}}$ cross section measurements in CMS}}
\vfill
\Author{Carmen Diez Pardos on behalf of the CMS Collaboration}
\Address{\institute}
\vfill
\begin{Abstract}
An overview of the most recent measurements of inclusive top quark pair production cross section is presented. The results are obtained using data collected with the CMS experiment in proton-proton collisions at centre-of-mass energies of 5.02, 7, 8 and 13 TeV. The data are compared with the predictions from perturbative QCD calculations at full NNLO+NNLL accuracy. 
\end{Abstract}
\vfill
\begin{Presented}
$9^{th}$ International Workshop on Top Quark Physics\\
Olomouc, Czech Republic,  September 19--23, 2016
\end{Presented}
\vfill
\end{titlepage}
\def\thefootnote{\fnsymbol{footnote}}
\setcounter{footnote}{0}

\section{Introduction}

The top quark is the heaviest known elementary particle and the only quark that decays before hadronisation, and thus gives direct access to its properties. With its large mass, it plays a crucial role in electroweak loop corrections, providing indirect
constraints on the mass of the Higgs boson. Top quark measurements also provide important input to QCD calculations. Moreover, various scenarios of physics beyond the standard model (SM) expect the top quark to couple to new particles.

Top quarks are mostly produced in pairs (${\rm t\bar{t}}$) via the strong interaction in hadron colliders. At LHC energies, the dominant mechanism is gluon-gluon fusion, corresponding to $\sim$80\% of the generation process. The top quark decays almost exclusively into a W boson and a b quark and it is the decay of the W bosons what defines the final state. Therefore, ${\rm t\bar{t}}$ signatures can be classified according to the combinatorics of the W boson decay: ${\rm t\bar{t}}$ final states include events with two leptons, two neutrinos and two b jets (dilepton), with one lepton, one neutrino and four jets, out of which two arise from a b quark (lepton+jets) or with six jets, out of which two stem from a b quark (fully hadronic channel). Only the most recent results in the three decay channels by the CMS~\cite{CMSColl} experiment are discussed. They are performed with the data sets collected at 7 TeV (5 fb$^{-1}$) and 8 TeV (19.8 fb$^{-1}$), the data collected in 2015 at 13 TeV (approximately 2.5 fb$^{-1}$) and 5.02 TeV (26 pb$^{-1}$).

\section{Dilepton (e$\mu$) decay channel}
\begin{figure}[htb]
\centering
\includegraphics[width=0.95\textwidth]{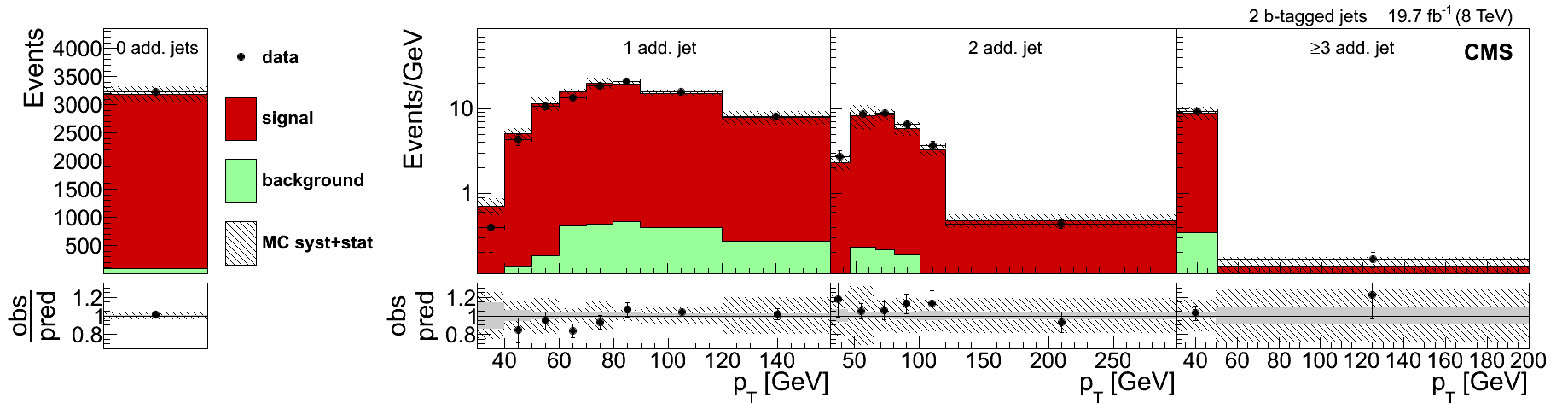}
\caption{Fitted event yields for events with two b-tagged jets at 8 TeV and for zero additional non-b-tagged jets (left) and p$_{\rm T}$ of the non-b-tagged jet with the lowest p$_{\rm T}$ for events with one, two, and at least three additional non-b-tagged jets (right)~\cite{top13004}.}
\label{fig:leptons}
\end{figure}

The most recent measurements in the dilepton decay channel require only events which contain an opposite charge electron-muon pair (e$\mu$). The restriction to the e$\mu$ channel allows obtaining a particular clean ${\rm t\bar{t}}$ event sample. At 7 and 8 TeV~\cite{top13004} no minimum number of jets or jets identified as stemming from the hadronization of a b quarks (b-tagged jets) is required. The cross sections are extracted with a simultaneous binned likelihood fit to the transverse momentum (p$_{\rm T}$) distribution of the non-b-tagged jet with the lowest p$_{\rm T}$ in the event, using different event categories of number of jets, b-tagged jets and lepton charge. An example of the post-fit distributions at 8 TeV is presented in Fig.~\ref{fig:leptons}. The cross sections result:
\begin{itemize}
\item $\sigma_{\rm t\bar{t}} = 173.6 \pm 2.1\, ({\rm stat}) \pm^{4.5}_{4.0} ({\rm syst})\pm 3.8\, ({\rm lumi})$ pb at $\sqrt{s}=7$ TeV,
\item $\sigma_{\rm t\bar{t}} = 244.9 \pm 1.4\, ({\rm stat}) \pm^{6.3}_{5.5}({\rm syst})\pm 6.4\, ({\rm lumi})$ pb at $\sqrt{s}=8$ TeV.
\end{itemize}
The precision is about 3.7\%.The main uncertainty sources of these measurements are luminosity, trigger and lepton identification. The ratio of the cross sections for the two different values of $\sqrt{s}$ results $R_{\rm t\bar{t}}=\sigma_{\rm t\bar{t}} (8\, {\rm TeV})/ \sigma_{\rm t\bar{t}} (7\, {\rm TeV})= 1.41 \pm 0.06$. The cross sections in the visible region, defined by the acceptance requirements on the two charged leptons in the final state (p$_{\rm T}>$20~GeV and $|\eta|<$ 2.4) are:
\begin{itemize}
\item $\sigma^{vis}_{\rm t\bar{t}}=  3.03 \pm 0.04\, ({\rm stat}) \pm^{0.08}_{0.07}({\rm syst})\pm 0.07\, ({\rm lumi})$ pb at $\sqrt{s}=7$ TeV and 
\item $\sigma^{vis}_{\rm t\bar{t}}=  4.23 \pm 0.02\, ({\rm stat}) \pm^{0.11}_{0.09}({\rm syst})\pm 0.11\, ({\rm lumi})$ pb at $\sqrt{s}=8$ TeV.
\end{itemize}

\begin{figure}[htb]
\centering
\includegraphics[width=0.35\textwidth]{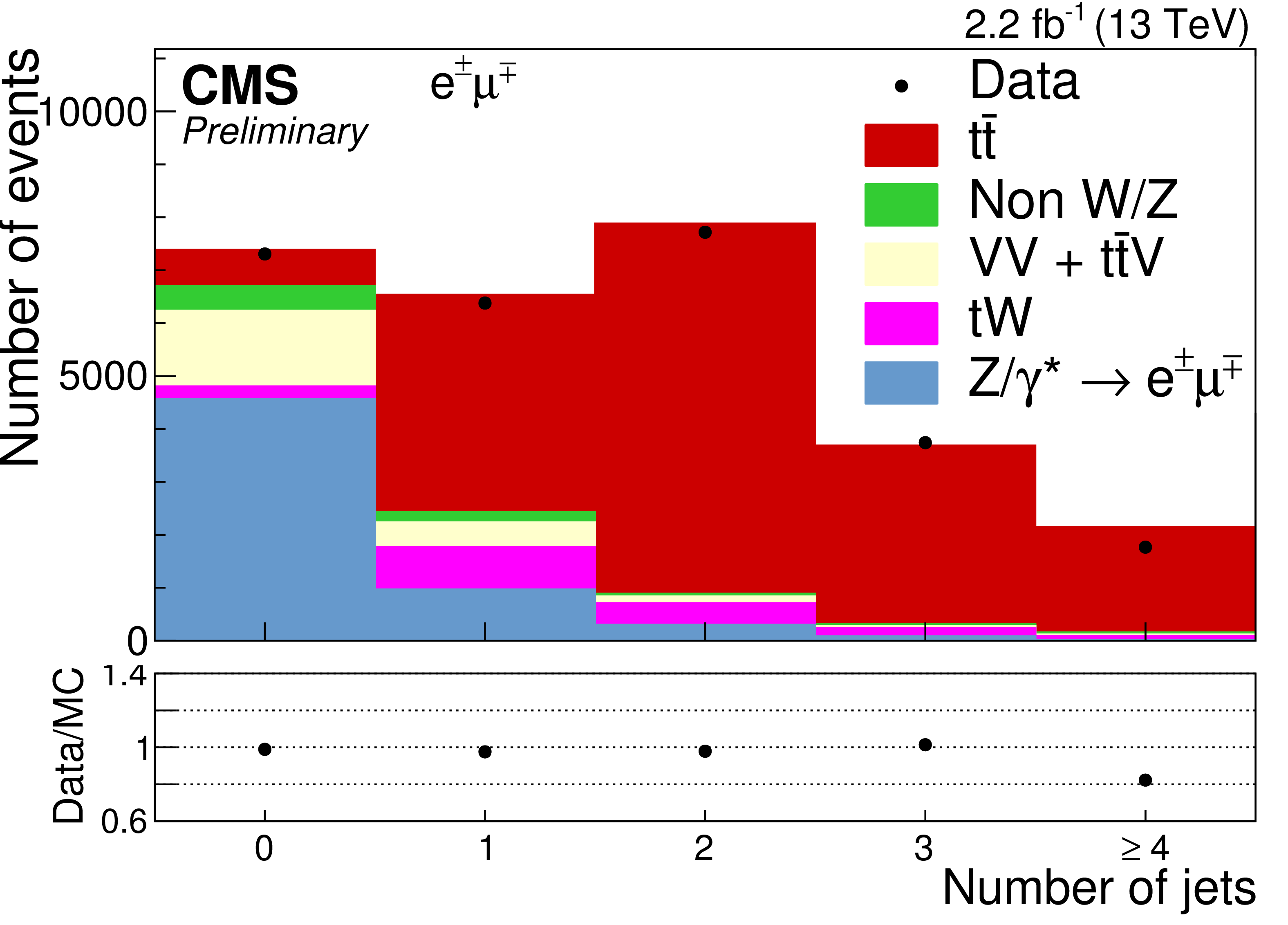}%
\includegraphics[width=0.35\textwidth]{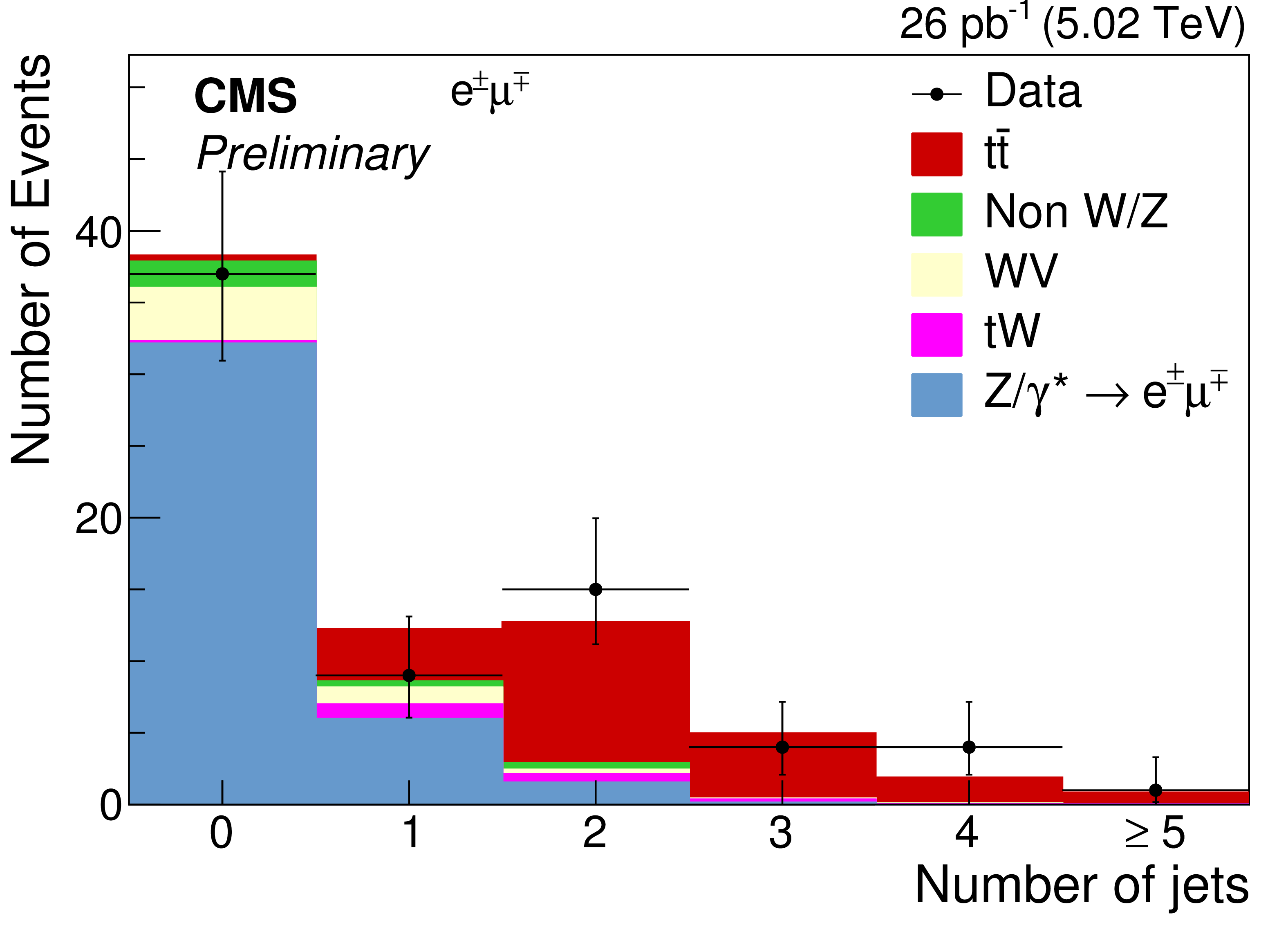}
\caption{Jet multiplicity distributions at 13 TeV (left)~\cite{top16005} and 5.02 TeV (right)~\cite{top16015}.}
\label{fig:leptons2}
\end{figure}

So far the approach used to perform the measurement with the data collected at 13 TeV~\cite{top16005} and at 5.02 TeV~\cite{top16005} is an event counting method. Both measurements select events with a least two jets. Additionally, the former requires the presence of at least one b-tagged jet. The jet multiplicity distributions for both $\sqrt{s}$ are shown in Fig.~\ref{fig:leptons2}. The background contributions arising from Drell-Yan and events with non-prompt leptons are estimated from data, while the contributions from single top quark, diboson events and $\rm{t\bar{t}}$ in association with bosons are taken from simulation. The result at 13 TeV is $\sigma_{\rm t\bar{t}}=793\pm 8 \, ({\rm stat})\pm38\, ({\rm syst})\pm 21\, ({\rm lumi})$~pb. The precision corresponds to 5.6\% and is limited by the uncertainty in the luminosity, lepton efficiencies, jet energy scale and signal modelling. The result at 5.02 TeV, $\sigma_{\rm t\bar{t}}=82\pm 20\, ({\rm stat}) \pm 5\, ({\rm syst})\pm 10\, ({\rm lumi})$~pb, is dominated by the statistical uncertainty (25\%).

\section{Lepton+jets decay channels}
\begin{figure}[htb]
\centering
\includegraphics[width=0.33\textwidth]{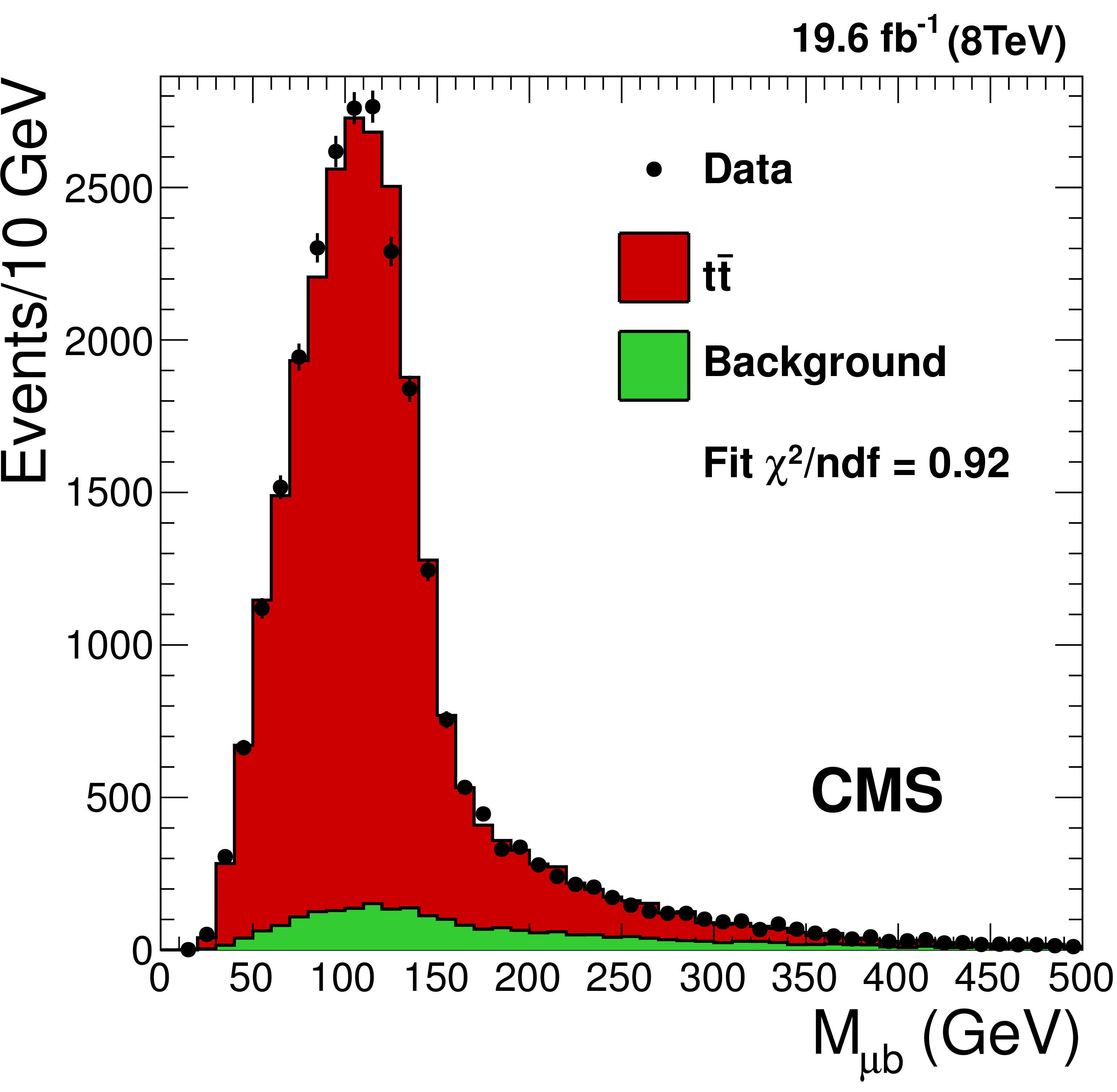}
\includegraphics[width=0.66\textwidth]{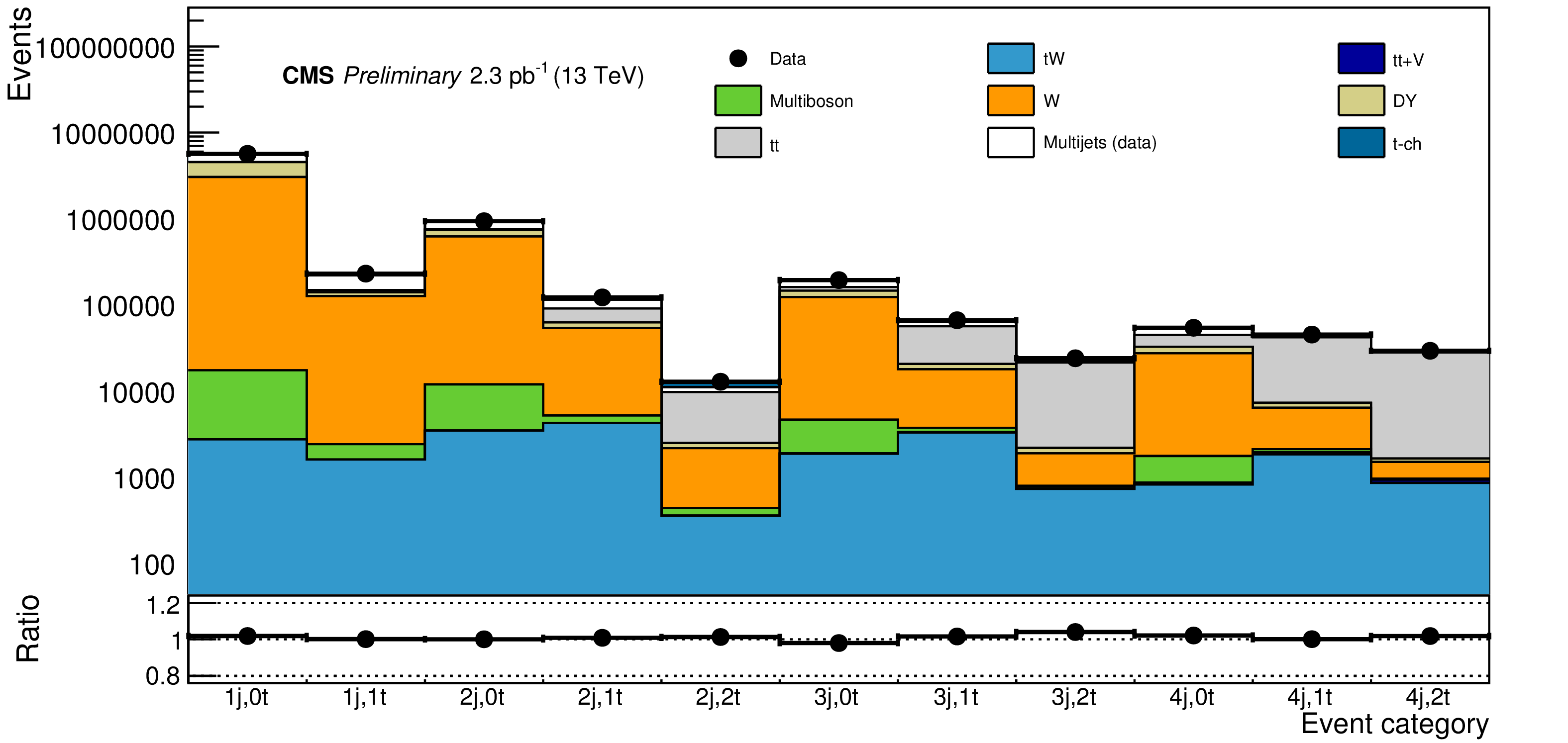}
\label{fig:ljets}
\caption{Distribution of M$_{{\rm lb}}$ after the fit at 8 TeV in the $\mu$+jets channel (left)~\cite{top12006}. Event yields for data and expected signal and background events for each of the categories of jet (j) and b-tagged jet (t) multiplicity (right)~\cite{top16006}.}
\end{figure}

The measurements in these channels at 7 and 8 TeV~\cite{top12006} select events that contain only one isolated electron or muon, at least four jets and at least one b-tagged jet. The method used to measure the cross section is a binned log likelihood fit to the invariant mass of the b-jet and the lepton (M$_{{\rm lb}}$), see Fig.~\ref{fig:ljets}. The calibration of the jet energy scale and the efficiency of b jet identification are determined from data. The cross section at 8 TeV results $228.9 \pm 3.4\, ({\rm stat}) \pm 13.7\, ({\rm syst}) \pm 6.0\, ({\rm lumi})$~pb. The main sources of systematic uncertainties are luminosity, jet energy scale and the hard scattering scale. The ratio of the cross sections at 8 TeV and 7 TeV is $R_{t\bar{t}}=1.43 \pm 0.04\,  ({\rm stat}) \pm 0.07\, ({\rm syst}) \pm 0.05\, ({\rm lumi})$, in very good agreement with the result in the e$\mu$ channel. The measurement at particle level in the visible phase space defined by the presence of exactly one muon or electron with p$_{\rm T}>$32~GeV and $|\eta|<$ 2.1, one neutrino with p$_{\rm T}>$40 GeV, and at least four jets with p$_{\rm T}>$40~GeV yields $\sigma^{vis}_{\rm t\bar{t}} = 3.80 \pm 0.06 \, ({\rm stat}) \pm 0.18 \, ({\rm syst})\pm 0.10\, ({\rm lumi})$ pb at $\sqrt{s}=8$ TeV. The inclusive cross section is also measured in the boosted regime, defined by the requirement that the top quark p$_{\rm T}$ is larger than 400 GeV. The measurement yields $\sigma_{\rm t\bar{t}} = 1.44 \pm 0.10 \, ({\rm stat+syst}) \pm 0.29\,  ({\rm theory}) \pm 0.04 \, ({\rm lumi})$ pb~\cite{boosted8TeV}. 
The measurement at 13 TeV~\cite{top16006} selects final states also with an isolated electron or muon, but only requires the presence of at least one jet. The events are divided in categories according to the number of jets and b-tagged jets. The cross section is obtained from a likelihood
fit to the M$_{{\rm lb}}$ distribution for each category. The cross section results $\sigma_{\rm t\bar{t}} = 834.7 \pm 2.5\, ({\rm stat}) \pm 20.7\, ({\rm syst}) \pm 22.6\, ({\rm lumi}) \pm 12.5 \,({\rm extrapol})$, which corresponds to a precision of about 4\%, limited by the uncertainty in the W+jets estimate, signal modelling uncertainties and luminosity. 

\section{Fully hadronic decay channel}
\begin{figure}[htb]
\centering
\includegraphics[width=0.35\textwidth]{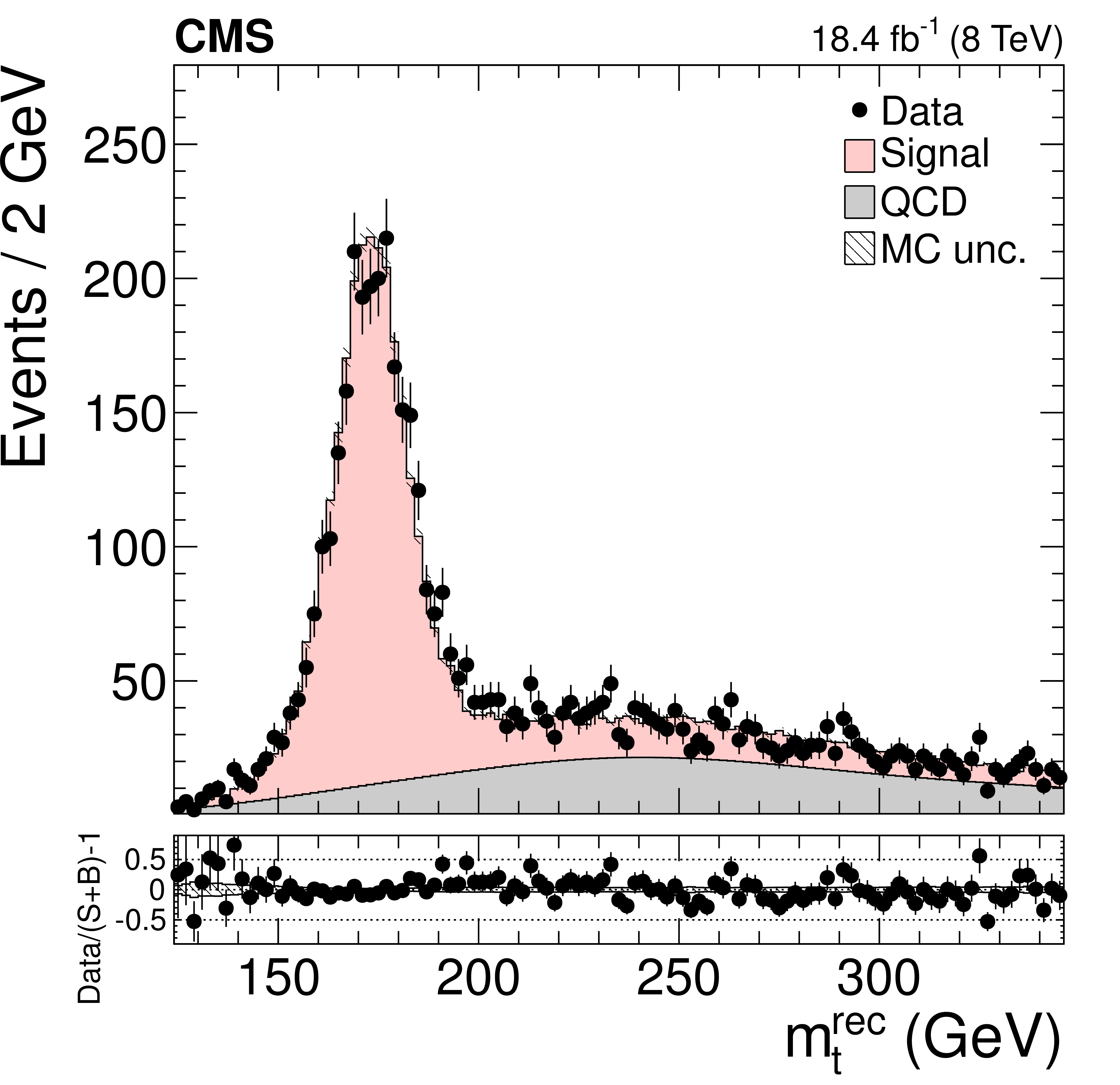}
\includegraphics[width=0.35\textwidth]{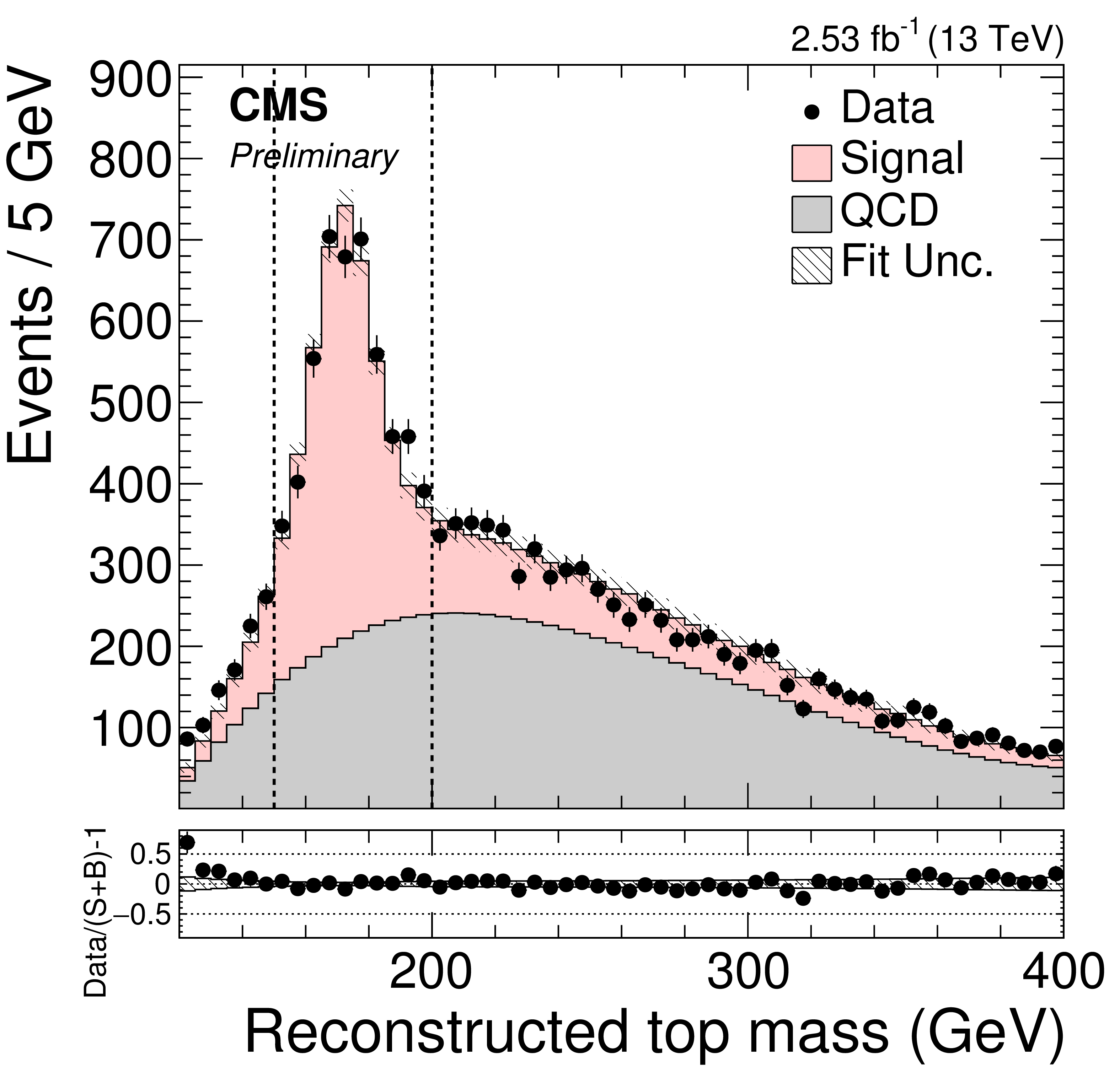}
\caption{Distribution of the reconstructed top quark mass at 8 TeV (left)~\cite{top14018} and 13 TeV (right)~\cite{top16013}.}
\label{fig:fullhad}
\end{figure}

The measurements are performed at 8 TeV~\cite{top14018} and 13 TeV~\cite{top16013}. Events in the fully hadronic decay channel are selected if they contain at least six jets, out of which two should be b-tagged. The dominant background contribution corresponds to QCD multijet events. To carry out the measurement, the $t\bar{t}$ system is first reconstructed. An unbinned maximum likelihood fit is then performed to the reconstructed top quark mass (m$_{\rm t}$) distribution to extract the normalization of the signal and background processes. The distributions at 8 and 13 TeV of m$_{\rm t}$ after the fit are shown in Fig.~\ref{fig:fullhad}. The dominant sources of systematic uncertainties are jet energy scale and b tagging efficiency. The measured cross sections are:
\begin{itemize}
\item $\sigma_{\rm t\bar{t}}=275.6 \pm 6.1\, ({\rm stat})\pm37.8\,({\rm syst})\pm 7.2\,({\rm lumi})$~pb at 8 TeV and
\item $\sigma_{\rm t\bar{t}}=834 \pm 25\, ({\rm stat})^{+118}_{-104}({\rm syst})\pm 23\,({\rm lumi})$~pb at 13 TeV,
\end{itemize}
with a precision of these of about 14\%. The preliminary result at 13 TeV also measures the cross section in the boosted regime (p$_{\rm T}^{\rm t}>400$~GeV). The result extrapolated to the full phase space of the ${\rm t\bar{t}}$ system is $\sigma_{\rm t\bar{t}}=727\pm 46\, ({\rm stat})^{+115}_{-112}({\rm syst})\pm 8\, ({\rm lumi})$~pb.

\section{Summary}
\begin{figure}[htb]
\centering
\includegraphics[width=0.33\textwidth]{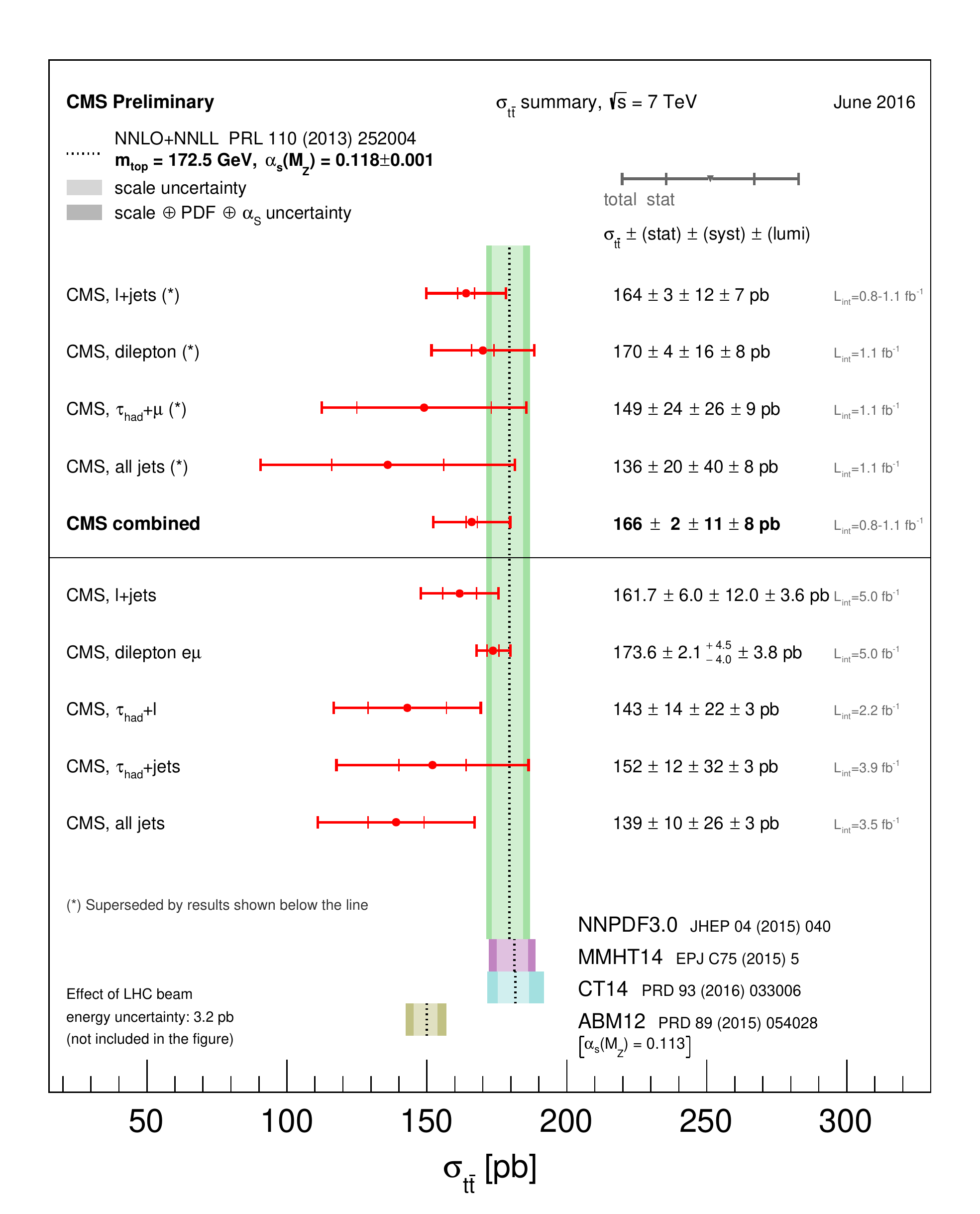}%
\includegraphics[width=0.33\textwidth]{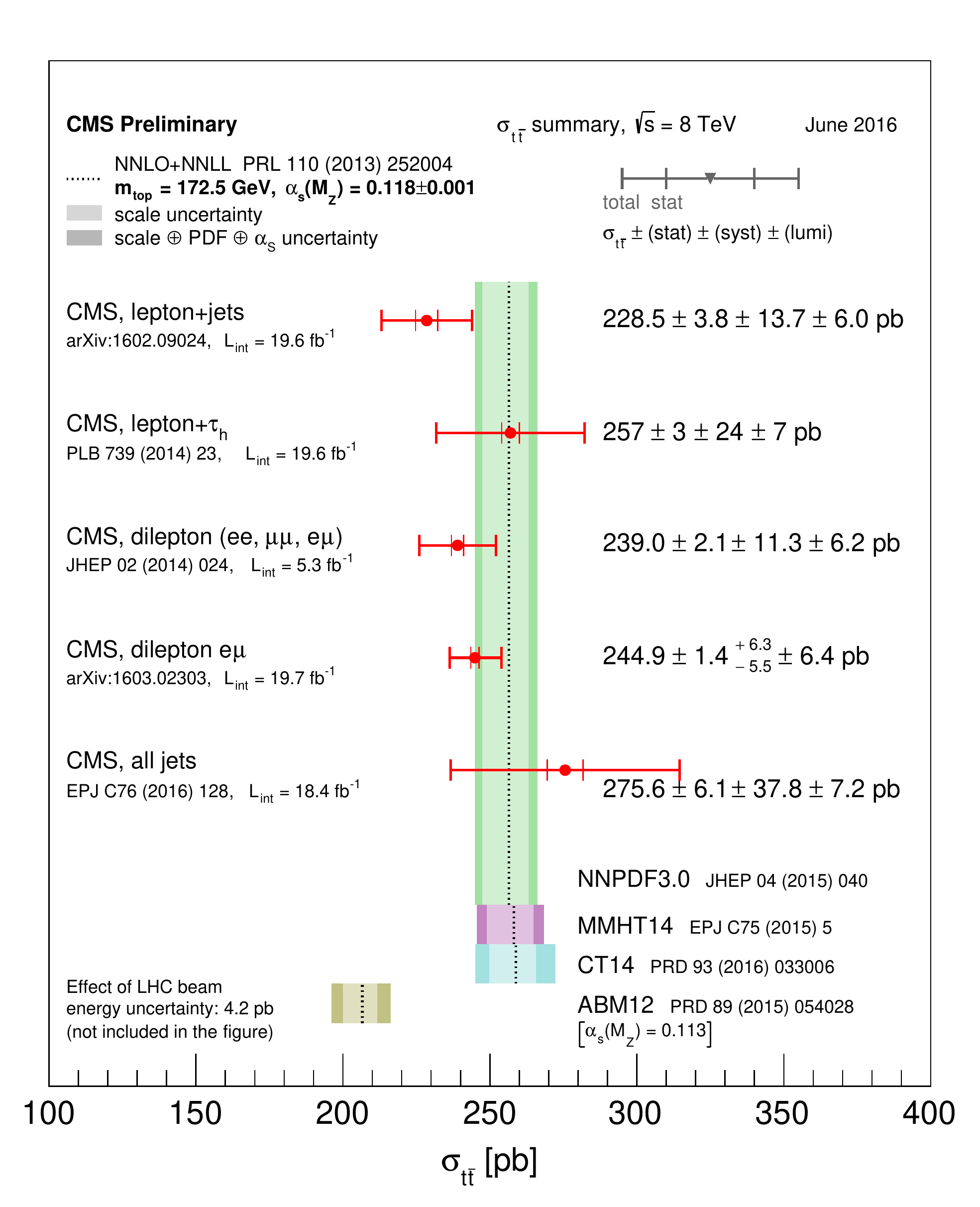}%
\includegraphics[width=0.33\textwidth]{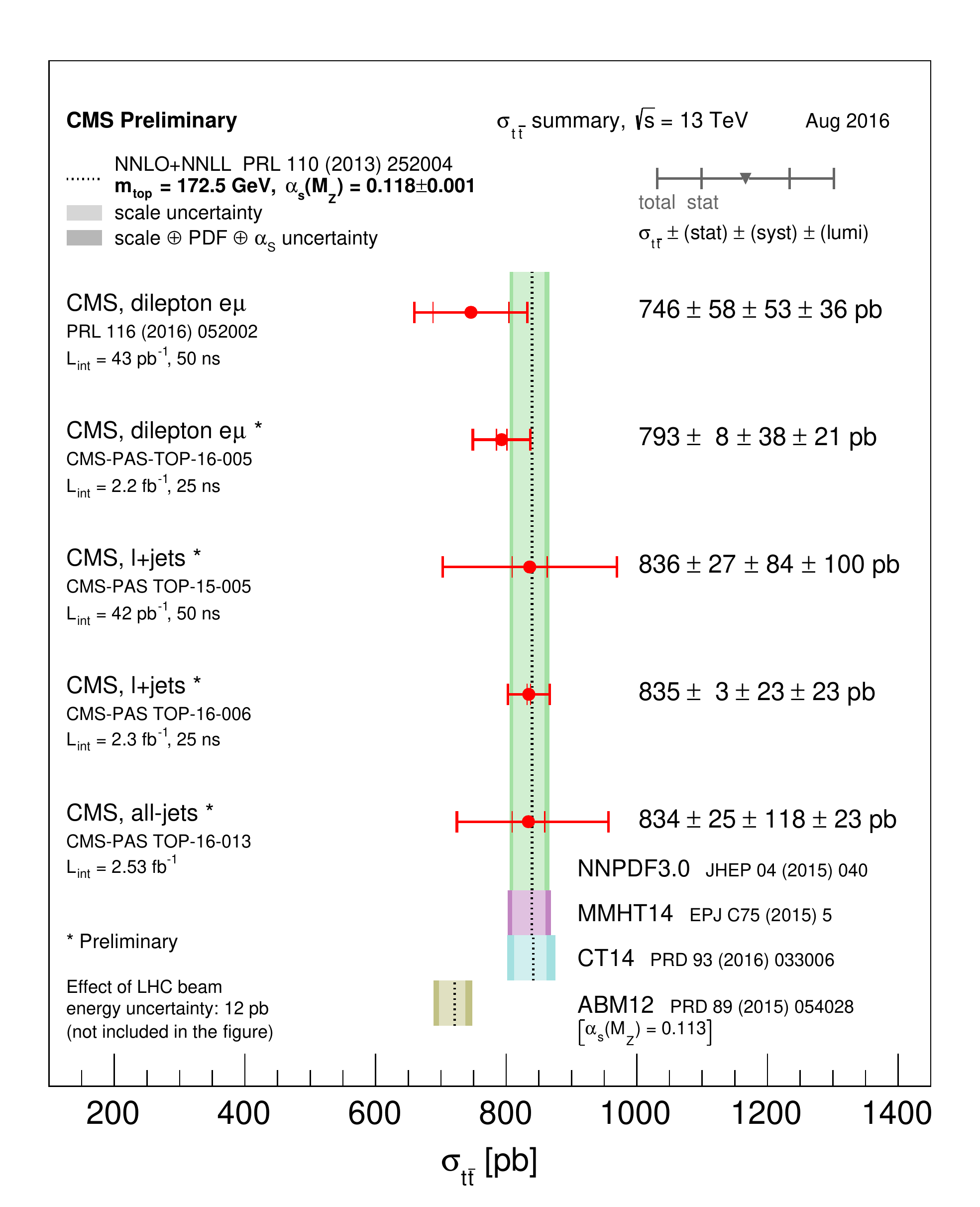}
\caption{Summary of the recent ${\rm t\bar{t}}$ measurements at 7 TeV (left), 8 TeV (middle) and 13 TeV (right) compared to the NNLO+NNLL predictions~\cite{sumplots}.}
\label{fig:summary}
\end{figure}

Top quark measurements provide important information about the production process as described in QCD, as well as sensitivity to possible new physics. In the last years, the LHC has become a real top factory and the large ${\rm t\bar{t}}$ data samples allowed to reach the high precision regime, below 4\%, of the same order of the precision of the full NNLO calculation. The results are in excellent agreement between the different channels and with the NNLO+NNLL predictions, see Fig.~\ref{fig:summary}.

\end{document}